\documentclass[prd,12pt,showkeys]{revtex4-1} 

\usepackage[latin1]{inputenc} 
\usepackage{amsmath, amsthm, amssymb, amsbsy,mathrsfs}
\usepackage{comment}
\usepackage{setspace}

\newcommand\fverb{\setbox\fverbbox=\hbox\bgroup\verb}
\newcommand\fverbdo{\egroup\medskip\noindent%
			\fbox{\unhbox\fverbbox}\ }
\newcommand\fverbit{\egroup\item[\fbox{\unhbox\fverbbox}]}
\newbox\fverbbox
\newcommand{\be}{\begin{equation}} 
\newcommand{\ee}{\end{equation}}
\newcommand{\lie}[1]{{\mathscr L}_{\raisebox{-1mm}{$\scriptstyle
      #1$}}}

\begin{document}

\title{Holography as Cutoff: a proposal for measure of inflationary
universes}

\author{Fábio Novaes\footnote{\footnotesize fabiomnsantos@df.ufpe.br} and}
\author{Bruno Carneiro da Cunha\footnote{\footnotesize bcunha@df.ufpe.br}}
\affiliation{Departamento de Física, Universidade Federal de Pernambuco,
53901-970, Recife, Pernambuco, Brazil}


\preprint{\today}	

\begin{abstract}
We propose the holographic principle as a dynamical cutoff for
any quantum theory of gravity, incorporating ideas of effective field
theory. This is done by viewing the holographic bound as a limit on
the number of degrees of freedom that can be turned on before the
geometrical picture of gravity loses applicability. We illustrate the
proposal by revisiting the problem of defining a measure for
homogeneous and isotropic spacetimes coupled to a scalar field and
conclude by discussing the implications to the single scalar field
inflationary model.  
\end{abstract}


\maketitle  

\section{\label{sec:intro}Introduction}

The concilliation between Quantum Mechanics and gravity, in the guise
of General Relativity, is the most important open problem of
theoretical physics. One of the difficulties can be attributed to the
fact that the Einstein-Hilbert lagrangian is non-renormalizable. As
classical (or effective) theories, non-renormalizable field theories
are fine, but as quantum theories they are heavily dependent on the
cutoff, and a cutoff scheme that respects the considerable
symmetries of General Relativity is still lacking. If such scheme were
to be found, we would be effectively dealing with a quantum system of a
finite number of degrees of freedom and no such obstacle to the
definition of a quantum version of General Relativity would exist.

At any rate, this problem has been somewhat circumvented in the last two
decades: in a very deep sense, one should not think of General
Relativity as a fundamental theory. The effective microscopic degrees
of freedom are not geometric, and the metric only arises as an
effective, macroscopical, description. Trying to quantize the space of
metrics directly is not only ineffective, it is also not a true
description of quantum gravity. 

On the other hand, many aspects of semi-classical gravity can be well
modelled by General Relativity: the laws of Black Hole Thermodynamics
and several descriptions of quantum effects of fields in curved
backgrounds rely heavily on the geometrical interpretation of quantum
gravity. In this sense, the microscopic description doesn't help us
gain insight on the geometrical character of many quantities. The
epitome of such quantities being the Bekenstein-Hawking entropy. 

The number of degrees of freedom available as area is perhaps the most
impressive result of semi-classical gravity. Not only it applies to
all macroscopical black-holes, but also for cosmological settings
\cite{Hooft1993,Fischler1998,Bousso2002}. The
universality of the result has a multitude of heuristic explanations,
but so far the microcanonical descriptions are non-geometrical in
nature. This lack of intuition about the result led several proposals
to promote it to a fundamental principle behind gravity itself.

In this letter we will attempt to operationalize the notion of quantum
gravity by using a truncated version of General Relativity as a
quantum theory, and the Holographic Principle as a means of defining a
natural cutoff of the theory. We will illustrate the procedure
considering the measure of initial conditions for an inflationary
universe and spend some time on the problem of naturalness.

\section{\label{sec:holo}Holography}

The idea that the number of degrees of freedom in some region of space
is bounded by the area of its boundary is a natural consequence of the
laws of Thermodynamics of Black Holes. In its essence, the argument is
deceptively simple: in the absence of any other forces of nature, the
natural endpoint for the evolution of any system is a black-hole. As
the second law of thermodynamics tells us that in such evolution the
entropy cannot decrease, then it should reach its maximum value for
a black hole. 

Upon closer inspection, however, the complications begin to
unfold. The interpretation of entropy as a measure of the number of
degrees of freedom available to the system leaves us at first with the
confusing task of recognizing which degrees of freedom of the system the
Bekenstein-Hawking entropy is counting. Again, a heuristic argument hints
at the solution: imagine a gas of free particles with temperature $T$,
interacting only gravitationally. As there is no force counteracting
gravity, the system will undergo gravitational collapse for generic
initial conditions. During the process, as the gas becomes denser and
denser, an observer sitting far away from the gas actually measures the
temperature of the gas {\it reducing} to the Black Hole
temperature.

We can draw two conclusions from this example: i) the
Bekenstein-Hawking temperature is assigned to degrees of freedom other
than the kinetic or the (other) potentials between the system; and ii)
The Bekenstein-Hawking entropy is so much larger than the usual values
found in non-gravitational systems that in the end it acts as a ``heat
bath'', such that the non-gravitational degrees of freedom thermalize with
the Black Hole temperature. Were it the other way around, having a black
hole at the end of the process would violate the second law of
thermodynamics for a large class of initial conditions.

The question that can be posed now is what happens with the
description of the system as the ``non-gravitational'' entropy
increases and surpasses the ``gravitational'' entropy? From the point
of view of general relativity, such system is unstable, exactly like a
supercooled or superheated system. The dynamics of the system is no
longer given by General Relativity since the ``thermodynamical''
variables like the metric are no longer well defined. The description
of the end result is to be achieved with a microscopic theory, and
even though one can still make predictions on general grounds, like
Hawking radiation -- or the Maxwell construction of the van der Waals
fluid -- the fate of the process relies ultimately on this underlying
theory. Questions like unitarity of evolution and presumed information
loss can only be answered there.

One can draw from the generic conclusions summing up the arguments above:
i) the Bekenstein-Hawking entropy counts the number of degrees of freedom
of gravity itself, and ii) The entropy can also be seen as a bound on the
number of degrees of freedom of gravity that have geometrical
interpretation. In other words, it can be seen as a cutoff on the
number of degrees of freedom of classical gravity.

Since the nature of this cutoff is very different from the usual maximum
momentum value, let us try then to make the idea a little more
concrete. Consider a globally hyperbolic spacetime,
{\it  i.e.}, with a global time function such that its topology is
$\mathbb{R}\times \Sigma$, with $\Sigma$ being connected. Let
$a$ be the ``radius of the universe'', that is, an
observable of $\Sigma$ which serves as the scale factor. We will label
a state $|a\rangle$ defined on the Hilbert space of gravity with
definite value for the operator $\hat{a}$.  The proposal can be made,
explicitly, by defining the maximum number of non-geometrical degrees
of freedom ${\cal N}(a)$ associated with the aforementioned states as:
\begin{equation}
{\cal N}(a)=\sum_{i}\langle a,\{i\}|a,\{i\}\rangle =
\frac{1}{4}\left(\frac{a}{\ell_P}\right)^{D-2}
\frac{2\pi^{(D-1)/2}}{\Gamma((D-1)/2)},
\label{eq:cutoff}
\end{equation}
the last fraction being the area of the unit sphere. The Planck length
is $\ell_P$. The index of summation $\{i\}$ refer to states which
share the same value for $a$. These may refer to ``matter''
anisotropies, or some other type of internal degree of freedom. We are
of course supposing that the number of degrees of freedom and the area
can be independently measured. If the area cannot be measured
independently of the scale factor $a$, then the functional form above
can be better justified without ``quantum corrections''. At any rate,
higher order corrections to the entropy will not change the discussion.

This cutoff shows up as a natural regulator of the purposed quantum
theory of gravity. Rather than being an inconsequential device to keep
amplitudes finite, this serves as a fundamental parameter of the
theory, in the sense of the role of the cutoff in an effective field
theory.

The usefulness of \eqref{eq:cutoff} as a cutoff is somewhat impaired in
the general case, since $a$ is a non-local function on $\Sigma$,
presumably also a non-local function of the degrees of freedom,
assuming that there are local degrees of freedom that are also gauge
invariant. However, there is a way of decoupling $a$ in a generic
class of metrics, still wide enough to be of interest. We will
formulate this in the following.

\section{\label{sec:infla}Initial Data and Inflationary Evolution}

The proposal for the cutoff is not entirely technical, though. To
illustrate the argument, let us define the scale factor in a more
geometrical fashion, and review the role of the Hamiltonian formalism
of General Relativity. To start, consider metrics with zero lapse vector:
\be
ds^2=-N^2dt^2+h_{ij}dx^i dx^j,
\ee
The hamiltonian formulation of gravity essentially treats $h_{ij}$ as
dynamical variables. Their conjugate momenta have an interesting
interpretation.  Let us define the
{\it extrisinc curvature}:
\be
K_{ab}=\frac{1}{2}\lie{n}h_{ab}=\frac{1}{2N}\dot{h}_{ab}
\ee
where $n^a$ is the unit vector in the time direction. By simple
algebra, we can relate $K_{ab}$ to the covariant derivative of
$n^a$, given that $h_{ab}=g_{ab}+n_an_b$:
\be
\begin{aligned}
K_{ab}&=\frac{1}{2}\left[ n^c\nabla_c h_{ab} +
  h_{cb}\nabla_a n^c+h_{ac}\nabla_b n^c\right] \\
&=(n^c\nabla_cn_{(a})n_{b)}+\nabla_{(a}n_{b)}
=h_{(a}^ch_{b)}^d\nabla_{c}n_{d}
\end{aligned}
\ee
{\it i.e.}, the projection of the covariant derivative of $n^a$ to the
spacial slice. Writing the curvature scalar in terms of the extrisinc
curvature we have (see, for instance, (E.2.29) in \cite{Wald}):
\be
^{D}R={^{(D-1)}R}-K^2+K_{ab}K^{ab},
\ee
where $^{(D-1)}R$ is the curvature scalar associated with
$h_{ab}$. The conjugate momentum to $h_{ab}$ can be now computed,
using the volume element $\sqrt{-g}=N\sqrt{h}$:
\be
\pi^{ab}=\frac{\partial S_\text{EH}}{\partial
  h_{ab}}=\frac{\partial {\cal L}}{\partial
  \dot{h}_{ab}}=\sqrt{h}(K^{ab}-Kh^{ab}).
\ee
And, by a Legendre transformation, one computes the Hamiltonian density:
\be
{\cal H}=\pi^{ab}\dot{h}_{ab}-{\cal L}
=-\sqrt{h}N\left[{^{(D-1)}R}-h^{-1}\left(\pi^{ab}\pi_{ab}-\frac{1}{D-2}
\pi^2\right)+{\cal H}_{\rm matter}\right]
\ee
For a scalar field, for instance, ${\cal H}_{\rm
matter}=1/2[(D\phi)^2+h^{-1}\Pi^2+V(\phi)]$,
where $\Pi=\sqrt{h}N^{-1}\dot{\phi}$ is the momentum conjugate to
$\phi$. Let us keep the expression generic for the time being.

Now assume that the only
relevant term to the evolution is the expansion of the
metric. $K_{ab}$ (and then $\pi^{ab}$) is then related to the
expansion factor: 
\be
K_{ab}=\frac{1}{N}\theta h_{ab} \quad \therefore \quad
\pi^{ab}=-\frac{D-2}{N}\theta\,\sqrt{h} h^{ab}.
\label{eq:expfac}
\ee
More importantly, the natural variable becomes the volume element
$\sqrt{h}$, to which we find the conjugate momentum:
\be
\frac{\partial S_\text{EH}}{\partial \sqrt{h}}=\frac{\partial
    S_\text{EH}}{\partial h_{ab}}\frac{\partial h_{ab}}{\partial
\sqrt{h}} 
=\frac{2}{\sqrt{h}}h_{ab}\frac{\partial S_\text{EH}}{\partial
h_{ab}}=\frac{2}{\sqrt{h}}\pi
\ee
With these formulas we can write the constraint ${\cal H}=0$ in terms
of $\sqrt{h}$ and the matter fields, in a Hamilton-Jacobi form:
\be
^{(D-1)}R+\frac{1}{4(D-1)(D-2)} \left(\frac{\partial S_\text{EH}}{\partial
  \sqrt{h}}\right)^2
 + {\cal H}_{\text{matter}}=0, \label{eq:hjeq}
\ee
For the type of metrics in which $K_{ab}\propto h_{ab}$, one can define
the scale factor as $\sqrt{h}=a^{D-1}$ and then $\theta$ in
\eqref{eq:expfac} is essentially the expansion factor $\dot{a}/a$. The
expansion factor is then the momentum conjugate to the scale factor, and
\eqref{eq:hjeq} is nothing more than (half) of the FRW equations written
in a fancier language.

The equation \eqref{eq:hjeq} encodes the dynamics of spacetimes on
which the anisotropies and inhomogeneities have been smeared out. By
several different accounts, such conditions are believed to hold
during the very early stages of the evolution of the universe. The
full grasp of the quantum mechanical system whose classical limit is
\eqref{eq:hjeq} would help us answer the question whether the
homogeneity and isotropy of the universe perceived now is the result
of cleverly chosen initial conditions, or the generic endpoint of the
dynamics of the spacetime. 

Now, the Hamiltonian constraint \eqref{eq:hjeq} encodes the tension
between the existence of a phase of Cosmological Inflation and the notion
of naturalness of initial conditions. If we assume that the expansion
rate is positive and relatively constant, \eqref{eq:hjeq} tells us that
${\cal H}_\text{matter}$ has to remain also relatively constant during a
``long'' span of time, essentially a cosmological constant. Although
other mechanisms of Inflation exist, the notion of a state with
relatively well-defined value for the expansion factor is at odds with
a well-defined value of the scale factor, since, as seen above, these
quantities are canonically conjugate. We will focus on whether such a
state with specific characteristics is actually ordinary in the phase
space of General Relativity.

Such definition of ``ordinary'' really depends on a working definition
of quantum gravity, as stressed by
\cite{Hollands:2002xi,Hollands:2002yb}. There is also an issue of the
global vs. local points of view, in which the two questions ``what is the
probability of inflation?'' and ``what is the probability of a sizeable
universe have gone through inflation?'' have different answers. We will
take a local point of view, in which quantum observables are formulated
in terms of locally measurable quantitites. Heuristically, one could think
of the quantum system as the ``microcanonical'' in the sense of
statistical mechanics, in which there
are a multitude of degrees of freedom not accessible macroscopically. The word ``microcanonical'' is perhaps inappropriate: in truth, there is no ``ensemble'' to speak of, and the analogy runs better if one thinks of the quantum system as sampling a sizeable portion of the available phase space, as in ergodic theory. The
macroscopical, ``thermodynamic'' system is the one whose phase space is
parametrized by $h_{ab}$ and $\pi^{ab}$, that is, General Relativity. Now,
a number of proposals for the measure on the thermodynamic phase space
have been proposed
\cite{Carroll:2010aj,Gibbons1987736,Hawking1988789,Gibbons:2006pa}, and
the problem of naturalness of inflation has been considered. As such, the
final answer relies deeply in regularization one uses to define a measure
in the (thermodynamic) phase space. Our proposal is to use the
Holographic principle, in the guise of \eqref{eq:cutoff}, to study the
thermodynamic phase space defined by the ``first quantized'' version of
the Hamiltonian constraint \eqref{eq:hjeq}.

\section{\label{sec:WDW-FRW}The Wheeler-DeWitt-FRW equation}

Our objective in this section is to introduce the Wheeler-DeWitt (WDW)
equation of the Friedmann-Robertson-Walker (FRW) model to obtain a
semiclassical description of the gravitational degrees of freedom. The
word semiclassical here means that we are assuming that quantum
perturbations of the metric are small compared to fluctuations of the
matter field. That is what allow us to restrict our considerations
only to FRW spacetime, which sets our considerations to classical
cosmology.

The WDW equation is obtained simply by casting the canonical
quantization prescription into the Hamiltonian constraint of General
Relativity. One starts with the 4-D Einstein-Hilbert lagrangian
coupled to a scalar field $\phi$, especialized to the FRW metric with
lapse function $N$ and spatial slices of (normalized) curvature $\kappa$:
\begin{equation}\label{eq:phiFRWlagrangian}
L = -\frac{3}{N}a\dot{a}^{2} + 3N\kappa a +
\frac{1}{2N}a^{3}\dot{\phi}^{2} - Na^{3}V(\phi) \;.
\end{equation}
from which we define the conjugate momenta 
\begin{equation}\label{eq:conjugatemomenta}
p_{a} = -\frac{6}{N}a\dot{a}\quad ,\quad  p_{\phi} =
\frac{1}{N}a^{3}\dot\phi \quad , 
\end{equation}
and obtain the hamiltonian
\begin{equation}\label{eq:hamiltonian}
\mathcal{H} = N\left(-\frac{p_{a}^{2}}{12a} +
\frac{p_{\phi}^{2}}{2a^{3}} + a^{3}V(\phi ) - 3\kappa a \right)\;.
\end{equation}
The equation of motion for $N$ show us that \ref{eq:hamiltonian} is
actually a constraint given by
\begin{equation}\label{eq:hamiltonianconstraint}
H \equiv  -\frac{p_{a}^{2}}{12a} + \frac{p_{\phi}^{2}}{2a^{3}} + a^{3}V(\phi
) - 3\kappa a\; = 0.
\end{equation}
which is just $\mathcal{H}=0$ with $N=1$. It is interesting for our
purpose to change the variable $a$ to $\alpha \equiv \log a^{3}$ which
is \textit{essentially} the volume of a patch of comoving volume
one. Now, using that
\[
p_{\alpha} =
-\frac{2}{3}e^{\alpha}\dot\alpha 
\]
in equation
\eqref{eq:hamiltonianconstraint}, we have
\begin{equation}\label{eq:alphaconstraint}
-\frac{3}{4}p_{\alpha}^{2} + \frac{1}{2}p_{\phi}^{2} +
e^{2\alpha}\left(V(\phi ) - 3\kappa e^{-2\alpha /3}\right)\; = 0.
\end{equation}
Now, if we impose the canonical quantization rule
\begin{equation}\label{eq:canonicalquantization}
p_{\alpha} \rightarrow -i\partial_{\alpha} \quad , \quad p_{\phi}
\rightarrow -i\partial_{ \phi }
\end{equation} 
into \eqref{eq:alphaconstraint}, we obtain a Klein-Gordon-like equation
for gravitation given by
\begin{equation}\label{eq:wdwfrw}
\left[\frac{3}{4}\frac{\partial^{2}}{\partial \alpha^{2}} -
\frac{1}{2}\frac{\partial^{2}}{\partial \phi^{2}} + e^{2\alpha }
\left(2V(\phi) - 3\kappa e^{-2\alpha /3}
\right)\right]\Psi(\alpha ,\phi ) = 0 \;,
\end{equation}
which is the Wheeler-DeWitt equation of our model. We can simplify it
with the substitution $\alpha \rightarrow \sqrt{3/2}\,\alpha$ giving 
\begin{equation}
\left[\frac{\partial^{2}}{\partial \alpha^{2}} -
\frac{\partial^{2}}{\partial \phi^{2}} + 2e^{\sqrt{6}\alpha}\left( V(\phi)
-
3\kappa e^{-\sqrt{2/3}\alpha} \right) \right]\Psi(\alpha ,\phi ) = 0
\;,
\label{eq:WDW-FRW}
\end{equation}
which we shall call the WDW-FRW equation. Note that $\alpha$ plays the
role of time translation, or scale-space translation. This will play
an important role in the following. 

With the WDW-FRW equation we will formulate a semiclassical
description of the thermodynamical degrees of freedom of gravitation
defined in the configuration space $(\alpha ,\phi )$. This ``first
quantized'' version of quantum gravity suffers from all the usual
pathologies: i) the spectrum is not bounded from below, which makes
the identification of the vacuum very difficult; ii) this problem is
exacerbated by the lack of apparent symmetries of the equation --
identification of the vacuum state as the one respecting all the
symmetries of the equations of motion is hopeless; iii) Without
time-invariance, one cannot talk about conservation of the number of
quanta, then the need of ``second quantization'' arises, with all the
interpretational problems of creation and annihilation operators of
quanta ensuing. We will gleefully overlook all these complications and
take \eqref{eq:WDW-FRW} as it stands: defining a semiclassical dynamic
on the reduced phase-space parametrized by $\alpha,\phi$.

\section{Probability of Inflation}\label{sec:probofinflation}

With all the subtleties disregarded, the system defined by
\eqref{eq:WDW-FRW} can be thought of as the wave equation in
two-dimensional space-time with a complicated,
``time-dependent''\footnote{That is, $\alpha$-dependent or
  $\phi$-dependent, depending on the chosen signature.}
potential. We will in this section review the calculation of the
number of degrees of freedom of the theory as the coincident limit of
a suitable Green's function of the system. Such Green's function will
satisfy the non-homogeneous equation:
\begin{equation}
\left[\frac{\partial^{2}}{\partial \alpha^{2}} -
\frac{\partial^{2}}{\partial \phi^{2}} + 2e^{\sqrt{6}\alpha}\left( V(\phi)
-
3\kappa e^{-\sqrt{2/3}\alpha} \right) \right]G(\alpha
,\phi;\alpha',\phi') = 
\delta(\alpha,\alpha')\delta(\phi,\phi'),
\label{eq:WDW-FRW1}
\end{equation}
which illustrates the ambiguity of such function: not only it is defined
up to the addition of a homogeneous term, but also in relativistic
theories it defines different Green's functions depending on the
prescription of contourning the poles. Moreover, whichever function
is picked will have an additional problem: in practice the coincidence
limit is divergent and needs to be regularized. 

In the following, we will argue that the right choice for Green's
function is the Hadamard function, and we shall also review the
point-splitting regularization method and show how to uniquely identify
the divergent nature of the Hadamard Green's function. Now we turn to the
definition of the density of states.

A standard procedure to count states in a quantum theory is conveniently
reproduced by analysing the spectral decomposition of the second order
hyperbolic equation
\begin{equation}\label{eq:formalWDW}
\left[\partial_{\alpha}^{2} + \hat K(\alpha ,\phi ) \right]\Psi =0
\end{equation}
where $\hat K(\alpha ,\phi)$ is a second order elliptic differential
operator. In particular, for the WDW-FRW model,
\begin{equation}\label{eq:ellipticWDW}
\hat K(\alpha ,\phi) = -\partial_{\phi}^{2} + 2e^{\sqrt{6}\alpha}\left(
V(\phi) - 3\kappa e^{-\sqrt{2/3}\alpha} \right).
\end{equation}
Unfortunately, equation \eqref{eq:formalWDW} is not separable. This
makes the analysis of the spectrum quite complicated because one
cannot directly Fourier transform the equation, reducing the problem
to an eigenvalue equation. However, there is another way to count
states by making an analytic continuation of $G_{W}$ and using the
Hadamard expansion.

\newcommand{\W}{\scriptscriptstyle W}
\newcommand{\alphap}{\alpha^{\prime}}
\newcommand{\phip}{\phi^{\prime}}
\newcommand{\x}{\underline x}
\newcommand{\y}{\underline y}

Let us recast \eqref{eq:WDW-FRW1} in a more convenient way by
factoring it with respect to the function $W(\alpha ,\phi)/\mu^{2}$
\begin{equation}\label{eq:modifiedwdw}
(\Box_{\W} + \mu^{2})G_{W}(\alpha ,\phi) = \frac{\delta (\alpha)
\delta (\phi )}{\sqrt{g_{\W}}}\;,
\end{equation}
where $G_{W}$ is the Green function of the WDW-FRW equation, the
function $W$ is given by
\begin{equation}\label{eq:supermetric}
W(\alpha , \phi) = 2e^{\sqrt{6}\alpha}\left( V(\phi) - 3\kappa
e^{-\sqrt{2/3}\alpha}
\right)\;,
\end{equation}
and $\mu$ is intended to be an analogue of the Klein-Gordon mass. In
this trick, we are interchanging a description of WDW-FRW in Minkowski
flat space with ``variable mass'' for a curved space version with mass
$\mu^{2}$ and metric
\[
ds^{2} = \frac{W(\alpha,\phi)}{\mu^2}(d\alpha^{2} - d\phi^{2})\;.
\]
We will be considering the case where $W>0$, so that the dynamics is
dominated by the interaction with matter. In the case where $W<0$,
$\alpha$ and $\phi$ will change roles.

$G_{W}$ is clearly not unique and depends on the boundary conditions,
being a fundamental solution of a partial diferential equation. On the
other hand, $G_{W}$ presents a singular structure in the coincident
limit $\alpha , \phi \rightarrow 0$ which is unique and is called the
Hadamard series expansion of $G_{W}$
\cite{Hadamard1923,Bunch:1978aq,Fulling1989}.
The Green's function may be formally written as the inverse operator
\[
G_{\W} = (\Box_{\W} + \mu^{2})^{-1} = i
\int_{0}^{\infty}ds\;e^{-i(\Box_{\W} + \mu^{2})s}\;,
\]
were an implicit $\epsilon$-prescription is supposed to define it as
a Feynman propagator.  Hence, setting $\underline x \equiv (\alpha
,\phi )$ and $\underline y \equiv (\alphap ,\phip )$, we obtain the
representation
\begin{align}\label{eq:greenint}
G_{\W}(\x,\y) &= i\int_{0}^{\infty}ds\, \langle\x\,\lvert e^{-is\Box_{\W}}
\rvert\,\y \rangle\, e^{-i\mu^{2}s}\nonumber\\
&\equiv i\int_{0}^{\infty}ds\, U(s;\x,\y)\, e^{-i\mu^{2}s}\;,
\end{align}
where $U(s;\x,\y)$ is the kernel of the associated Schrödinger equation
\[
i\frac{\partial U}{\partial  s} = \Box_{\W}U\;.
\]
In order to solve the initial value problem, $U$ must obey the
boundary condition
\begin{equation}\label{eq:boundarycond}
 \lim_{s\rightarrow 0}\, U(s;\x,\y) = \frac{\delta \left(\x -
\y\right)}{\sqrt{g_{\W}}}\;.
\end{equation}

There is an easy way to see that the coincidence limit of $G_{W}(\x
,\y)$ gives a formula for the number of states. Consider that the
Hilbert space of solutions of \eqref{eq:modifiedwdw} is labeled by the
parameter $g$. The number of states is thus given by
\begin{align*}
\mathcal{N} &\equiv  \sum_{g}\, 1 \\
&= \sum_{g} \langle0;g\rvert\, \delta(\Box_{W}+\mu^{2})\, \rvert 0;g\rangle \\
&= \frac{1}{\pi }\sum_{g} \langle0;g\lvert\,\text{Im}\,\left(i
\int_{0}^{\infty } ds\, e^{-i(\Box_{W}+\mu^{2})s}
\right)\rvert 0;g\rangle \\
&= \frac{1}{\pi}\sum_{g} 
\int d^2x\; |W|\,\text{Im}\, G_{W}(\x ,\y)\,\psi_{g}^{*}(\x)\psi_{g}(\y)\\
&= \frac{1}{\pi}\int dx \;|W|\,\text{Im}\, G_{W}(\x ,\x)\;,
\end{align*}
where the delta function in the second line must be understood as a
projection operator and in the fourth line we used the completeness
relation
\[
\sum_{g} \psi_{g}^{*}(\x)\psi_{g}(\y) = \delta (\x -\y )\;.
\]
Now to properly define the number of states above we will use the
point-splitting method to regularize $G_{W}$.

First we note that eq.~\eqref{eq:boundarycond} sets the singularity
structure of $U$ with respect to $s$ and an appropriate
\textit{Ansatz} for a $d$-dimensional manifold in terms of the
geodesic distance $\sigma$ is
\[
U(s; \x ,\y) = \Delta^{\scriptscriptstyle 1/2}(\x,\y)\; \frac{1}{(4\pi
i s)^{d/2}}\;e^{i\sigma (\x ,\y)/2s}\,\sum_{n=0}^{\infty} a_{n}(\x
,\y)(is)^{n}\;,
\]
which is just the VanVleck-Morette determinant
$\Delta^{\scriptscriptstyle 1/2}(\x,\y)$ times the product of a delta
sequence of functions (set to match the delta boundary condition) and
a smooth function of $s$
\cite{Dewitt1960,Christensen1978,Fulling1989}. Putting
this $U$ into \eqref{eq:greenint}, we obtain
\begin{align*}
G_{\W}(\x ,\y) &= -\frac{\Delta^{\scriptscriptstyle 1/2}}{(4\pi
i)^{d/2}}\int_{0}^{\infty}ds\; \frac{1}{s^{d/2}}\;e^{-i(\mu^{2}s
-\sigma/ 2s)}\,\sum_{n=0}^{\infty} a_{n}(is)^{n}\\
&= -\frac{\Delta^{\scriptscriptstyle 1/2}}{(4\pi
i)^{d/2}}\sum_{n=0}^{\infty}a_{n} \left(-\frac{\partial}{\partial
\mu^{2}}\right)^{n}\int_{0}^{\infty}ds\; \frac{1}{s^{d/2}}\;e^{-i(\mu^{2}s
-\sigma/ 2s)}  \;.
\end{align*}
The last equation involves an integral representation of the Hankel
function of the second kind and may be explicitly written for the case
$d=2$ as \cite{Bunch:1978aq}
\[
G_{W}(\x ,\y) = \frac{1}{4}\Delta^{\scriptscriptstyle 1/2}
\sum_{n=0}^{\infty}a_{n} \left(-\frac{\partial}{\partial
\mu^{2}}\right)^{n} H_{0}^{(2)}[(-2\mu^{2}\sigma)^{1/2}]\;.
\]
Using the resort of an asymptotic series expansion for $H_{0}^{(2)}$,
we obtain a series for $G$ and, noting that $G^{(1)} = 2\,\text{Im}G
$, we have \cite{Bunch:1978aq}
\begin{align}\label{}
 G^{(1)}(\x,\y) &= \Delta^{\scriptscriptstyle 1/2} \{-L[1 +
\frac{1}{2}\mu^{2}\sigma -\frac{1}{2}a_{1}\sigma  + {\cal
O}(\sigma^{2})]\\
&+ \frac{1}{2}[\mu^{-2}a_{1} + \mu^{-4}a_{2} + {\cal O}(\mu^{-6})]\\
&+ \frac{1}{4}\sigma [2\mu^{2} -a_{1} -\mu ^{-2}a_{2} + {\cal
O}(\mu^{-4})] +
{\cal O}(\sigma^{2})\}\;,
\end{align}
where $L= \gamma + \frac{1}{2}\ln\lvert\frac{1}{2}\mu^{2}\sigma
\rvert$. As our interest remains in the coincidence limit $\sigma =0 $
of $G$, we use the fact that the coefficients $a_{n}$ may be
calculated quite directly in this limit \cite{Christensen1978}, giving
\begin{align}\label{}
[\Delta^{1/2}] &= 1 \, ;\\
[a_{1}] &= \frac{R}{6}\, ;\\
[a_{2}] &= \frac{1}{60}(R^{2} + 2\Box R)\,.
\end{align}
As the mass $\mu$ has no physical meaning for us, we set $\mu = 1$ and
taking the limit $\sigma \rightarrow 0$ we see that the Hadamard
series of $G^{(1)}$ has the form
\begin{equation}\label{eq:singularity}
G^{(1)}(\sigma ) \,\sim\, A\ln\sigma + B
\end{equation}
and, hence, presents a logarithm singularity. According to the
asymptotic theory of partial differential equations
\eqref{eq:singularity} represents the general singular structure of
Green's functions of Klein-Gordon equation for $d=2$. We have to find
in this divergent expression a measure of the number of degrees of
freedom available to the field. Being of the hyperbolic type, the
Klein-Gordon equation allows for an infinite number of solutions, but
this infinite number comes about because we are including non-physical
excitations with arbitrarily small wavelength. We then need a
regularization scheme that reflects the physics of the problem; the
situation so far not unlike the problem in flat space. We then choose
to deal not with the terms in the expansion $A$ and $B$ in
\eqref{eq:singularity} directly but instead to study their dependence
on the factor $W$. Again, the situation is not unlike ordinary quantum
field theory in curved spaces where the choice of a vacuum in flat
space fixes the relative energy between vacuum in confined spaces
(Casimir effect), which are related to the Minkowski vacuum by a
conformal transformation. Coming back to \eqref{eq:singularity}, the
term $A$ has the interpretation of the normalization of the
wave-function, so it does not encode the geometric dependence of the
degrees of freedom. This leaves us with the term $B$, which should
encode the dependence of the number of degrees of freedom. This will
be the focus of our attention from now on.

Finally, for the Green's function of eq.~\eqref{eq:modifiedwdw},
$A=-1/2\pi$ and
\begin{equation}\label{eq:V}
B = \frac{1}{\pi}\left[-\gamma + \log\sqrt{2}+ \frac{R}{12} +
\frac{1}{120}(R^{2} + 2\Box R) +{\cal O}(R^4,\Box^2R)\right]\;.
\end{equation}
where $R$ is given by:
\begin{equation}
 R=-\frac{1}{2}\left(\frac{V''(\phi)(V(\phi)-3\kappa
e^{-\sqrt{2/3}\alpha})-(V'(\phi))^2+2\kappa
e^{-\sqrt{2/3}\alpha}V(\phi)}{e^{\sqrt{6}\alpha}(V(\phi)-3\kappa
e^{-\sqrt{2/3}\alpha})^3}\right)
\label{eq:curvaturescalar}
\end{equation}
One notes that $R$ becomes large in the regime $V(\phi)\approx 3\kappa
a^{-2}$, corresponding to configurations where the potential energy
rivals the curvature of the spatial slices. This should be considered a
true quantum regime where the semiclassical approximation does not hold.
Classically, the quantity $V(\phi)-3\kappa e^{-\sqrt{2/3}\alpha}$
corresponds to
the Hubble parameter squared in the FRW equations. We will then consider
the regime where this number is positive, in such a way that $R$ is
small. Note that this corresponds roughly with the regime where
the derivatives of the potential are much smaller than the potential
itself, a requisite for slow-roll inflation.

In the semiclassical limit, if the terms of higher order in $R$ are
neglected, we have
\[
\mathcal{N} = \frac{1}{2\pi^{2}} \int d\alpha d\phi\;
|W(\alpha , \phi)|\left[C + \frac{R}{12} \right]
\]
where $C=-\gamma + \log\sqrt{2}$. This is the formal number of states
of the Hilbert space of solutions of the WDW-FRW equation, in the
point-splitting regularization scheme. The measure is still infinite
after an integration over the whole $(\alpha ,\phi )$ plane. However,
we are mainly interested on the derivative of $\mathcal{N}$, that is,
$\mathcal{N}({\alpha })$, which represents the instantaneous number of
states for a given size of the universe $\alpha$. It should be noted
that the value of $C$ changes according to the regularization scheme
used. On the other hand, one can try to define a {\it relative} number
of available states by comparing the expression above with some
``fiducial'' expression. This can be argued as follows: in the absence
of a potential, $R=0$ according to \eqref{eq:curvaturescalar}. In this
situation, one has ``pure gravity'', and the number of degrees of
freedom has to express this fact. Scaling back $\alpha \rightarrow
\sqrt{2/3}\,\alpha$ and using the holographic bound to set the range of
values allowed for the scalar field:
\begin{equation}
 {\cal N}(\alpha)=\frac{3|\kappa|}{\pi^2}e^{4\alpha/3}|\Delta\phi|
\Longrightarrow |\Delta\phi|=Ke^{-2/3\alpha}
\label{eq:spreadphi}
\end{equation}
for some constant $K$. In this manner the number of states associated
with pure gravity scales as the area $e^{2\alpha/3}$. This can also be
seen as a ``gravitational uncertainty principle'', which states that the
spread on the values of the scalar field decreases with the inverse of
the square of the scale factor. Large universes would have a small spread
of the values of the scalar field, then. One should point out that the
interval specified by \eqref{eq:spreadphi} need not be centered in
$\phi=0$. In general there is no symmetry forcing this to be
the case. At this point it should be noted that the holographic bound
effectively serves as a cutoff on the number of physically available
solutions of the Klein-Gordon equation, in the sense that only the
number of states given by ${\cal N}(\alpha)$ have an interpretation as
a geometrical, macroscopic universe. This is successfully incorporated
in the quantum theory as a dynamical limit. It is also relevant that
this cutoff can take into account anisotropic and dynamical states, as
long as they are eigenstates of the scale factor operator
$\hat{a}$. In other words, instead of having only one local state
labelled with $\alpha$ and $\phi$ in \eqref{eq:modifiedwdw}, one can
have up to ${\cal N}(\alpha)$ states associated with these macroscopic
quantities. These presumably relating to anisotropic degrees of
freedom of these quantities.

Let us digress a little on the curvature terms. We note that, for
$\kappa=0$ in \eqref{eq:supermetric}:
\begin{align*}
R &= \frac{1}{W}\frac{d^{2}}{d\phi^{2}}\log V(\phi)\\
&= \frac{1}{W}\frac{d}{d\phi}\left(\frac{V^{\prime}}{V} \right)\\
&= \frac{1}{W}\,\frac{d}{d\phi}\sqrt{2\epsilon}\;,
\end{align*}
which allows to relate the Ricci scalar $R$ with the inflationary
slow-roll parameter $\epsilon = V^{\prime}/2V$. Obviously its powers and
derivatives are associated with the curvature corrections in
\eqref{eq:V}. If slow-roll inflation is to take place, {\it we then want
the curvature corrections to be small}.

Therefore, the total number of states for a given
size of the universe is then approximately given by the ``volume
term'':
\begin{equation}
\mathcal{N}(\alpha ) \approx \frac{1}{2\pi^2}\int d\phi\, |W(\alpha,\phi)|
\approx\frac{1}{\pi^{2}}e^{2\alpha}\, V(\phi_c)|\Delta\phi|\label{
eq:numberofstates }.
\end{equation}
In which we can state the problem: this number is also limited by the
holographic bound $\propto e^{2\alpha/3}$. Using the result
\eqref{eq:spreadphi}, we see that the most probable value of the scalar
field $\phi_c$ is then veered towards {\it decreasing} the value of
the potential $V(\phi_c)\propto e^{-2\alpha/3}$. This value, however,
is incompatible with inflation because of the Hamiltonian constraint:
$H^2\propto V(\phi)$ induces a linear growth of the scale-factor
$a\propto t$, where $H$ is the Hubble parameter.

One can conclude that, by this proposal, slow-roll inflation is supressed
by entropic grounds: the volume of phase space of initial conditions of
gravity that allow for inflation is just too small for it to be favored
dynamically. This conclusions, of course, need not apply to all different
models of inflation, which in itself makes for an interesting follow-up
to this discussion.

\section{Comments}

The main purpose of this letter is to introduce the concept of the
holographic bound as a cutoff for a theory of quantum gravity. The nature
of this cutoff is very similar to that of effective
``non-renormalizable'' field theories: it is a crucial part of the
theory, and dictates large-scale behavior, even though the cutoff belongs
to high-energy phenomena. 

We motivated the employment of the holographic bound seeing it not as
a ``true bound'' on the degrees of freedom, but rather as a limit
under which these degrees have geometrical interpretation. The picture
that arises is a bit different from usual field theories: rather than
a bound on the observables on the system (like the energy), we have a
non-local -- albeit in parameter space -- bound that counts all states
labelled by a single scale factor. Because of the technical issues,
the discussion was kept to spatially homogeneous FRW cosmology coupled
to a single scalar field. Using standard procedures to count the total
number of states, we could argue that usual slow-roll inflation is
entropically supressed with this provision.

While the particular conclusion for slow-roll inflation is model
dependent, the application of the holographic bound as a cutoff on the
number of degrees of freedom in quantum gravity can be used as an
universal guideline for studying semiclassical effects in quantum
gravity, beyond the Laws governing the temperature of the fields coupled
to classical gravity. 

The proposal put forward here can be contrasted to the many existent in
the literature as being both natural and coordinate independent. The
apparent drawback of only be applicable to spherically symmetric
situations does not hinder many of the applications, both in cosmology
and in black hole dynamics. This point opens an avenue of future
developments.

\section*{Acknowledgements}

We would like to thank Amílcar Queiroz, Luciano Barosi and Francisco
Brito for fruitful discussions. One of the authors (FN) acknowledge
partial support from CNPq. 

\bibliographystyle{hieeetr} 

\end{document}